\documentclass[twocolumn,aps,prd,superscriptaddress,nofootinbib]{revtex4-1}
\usepackage{graphicx}
\graphicspath{{figures/}{fig/}}
\DeclareGraphicsExtensions{.eps}
\usepackage{amsmath}
\usepackage{bm}
\usepackage{slashed}
\usepackage{epsfig}
\usepackage{amsfonts}
\usepackage{epstopdf}
\usepackage{color}
\usepackage{extarrows}
\newcommand{\ben}{\begin{eqnarray}}
\newcommand{\een}{\end{eqnarray}}

\newcommand{\bef}{\begin{figure}[!htp]}
\newcommand{\eef}{\end{figure}}

\newcommand{\bea}{\begin{eqnarray}}
\newcommand{\eea}{\end{eqnarray}}

\def\ba{\begin{linenomath*}\begin{equation}}
\def\ea{\end{equation}\end{linenomath*}}

\allowdisplaybreaks

\newcommand{\sect}[1]{\section{#1}}
\begin{document}
\title{Multiplicative renormalizability of quasi-parton operators}

\author{Zheng-Yang Li}
\email{lizhengyang@pku.edu.cn}
\affiliation{School of Physics and State Key Laboratory of Nuclear Physics and
	Technology, Peking University, Beijing 100871, China}
\author{Yan-Qing Ma}
\email{yqma@pku.edu.cn}
\affiliation{School of Physics and State Key Laboratory of Nuclear Physics and
	Technology, Peking University, Beijing 100871, China}
\affiliation{Center for High Energy physics, Peking University, Beijing 100871, China}
\affiliation{Collaborative Innovation Center of Quantum Matter,
	Beijing 100871, China}
\author{Jian-Wei Qiu}
\email{jqiu@jlab.org}
\affiliation{Theory Center, Jefferson Lab, 12000 Jefferson Avenue, Newport News, VA 23606, USA}

\date{\today}

\begin{abstract}
Extracting parton distribution functions (PDFs) from lattice QCD calculation of quasi-PDFs has been actively pursued in recent years. We extend our proof of the multiplicative renormalizability of quasi-quark operators in Ref.~\cite{Ishikawa:2017faj} to quasi-gluon operators, and demonstrated that quasi-gluon operators could be multiplicatively renormalized to all orders in perturbation theory, without mixing with other operators. We find that using a gauge-invariant UV regulator is essential for achieving this proof.  With the multiplicative renormalizability of both quasi-quark and quasi-gluon operators, and QCD collinear factorization of hadronic matrix elements of there operators into PDFs, extracting PDFs from lattice QCD calculated hadronic matrix elements of quasi-parton operators could have a solid theoretical foundation.
\end{abstract}
%\pacs{12.38.Bx, 13.88.+e, 12.39.-x, 12.39.St}

\maketitle
\allowdisplaybreaks

%%%%%%%%%%%%%%%%%%%%%%%%%%%%%%%%%%%%%%%%%%%%%%%%%
\sect{Introduction}	
\label{sec:intro}
The parton distribution functions (PDFs) encode important nonperturbative information of strong interactions. Based on QCD factorization \cite{Collins:1989gx}, PDFs have been successfully extracted from high-energy scattering data with a good precision \cite{Lin:2017snn}. However, from both theoretical and practical points of view, extracting PDFs from first principle lattice QCD (LQCD) calculations must be done for testing non-perturbative sector of QCD, as well as needed for studying partonic structure of hadrons that could be difficult to do scattering experiments with. 

Calculating PDFs in Euclidean-space LQCD {\it directly}, if not impossible, is difficult due to the time-dependence of the operators defining PDFs \cite{Lin:2017snn}.  A novel approach was suggested by Ji \cite{Ji:2013dva}, who introduced a set of LQCD-calculable quasi-PDFs and argued that the quasi-PDFs of hadron momentum $P_z$ become corresponding PDFs when $P_z$ is boosted to infinity. A number of other approaches to extract PDFs from LQCD calculations were also proposed~\cite{Liu:1993cv,Liu:1999ak,Liu:2016djw,Orginos:2017kos,Chambers:2017dov}. In Refs.~\cite{Ma:2014jla,Ma:2017pxb}, two of us proposed a QCD factorization based general approach to calculate PDFs in LQCD {\it indirectly}. 
Similar to the extraction of PDFs from experimental data of factorizable and measurable hadronic cross sections, we proposed to 
extract PDFs by the global analysis of data generated by LQCD calculations of good ``lattice cross sections" (LCSs), which are defined as hadronic matrix elements that satisfy (1) calculable in Euclidean-space LQCD, (2) renormalizable for ultraviolet (UV) divergences to ensure a reliable continue limit, and (3) factorizable to PDFs with infrared-safe matching coefficients. It is the (3)-factorization that relates the desired PDFs to the LQCD calculable LCSs. 

To extract the rich, precise and flavor separated information on PDFs, it is necessary to find as many good LCSs as possible, since different flavor PDFs are likely to contribute to the same LQCD calculated LCSs.  For constructing good LCSs, we studied two types of operators in terms of (a) correlation of two gauge-dependent field operators with proper gauge links, which we call quasi-parton operators  since they cover all operators defining quasi-PDFs and more, and (b) correlation of two gauge-invariant currents. In our approach, LQCD calculation of each of these good LCSs in coordinate space provides the needed information to {\it constrain} the PDFs, similar to the role of measuring various experimental cross sections to {\it constrain} the PDFs. For a comparison, in Ji's proposal \cite{Ji:2013dva}, one focuses on reproducing PDFs by corresponding quasi-PDFs calculated in LQCD at a large enough $P_z$.

For LQCD calculations, it is relatively less expensive to calculate the type (a) operators comparing with type (b) operators, while the renormalization of the type (b) operators is much more simpler than that of the type (a) operators.
Renormalization of the type (b) operators is almost trivial, for which one only needs to renormalize the gauge-invariant local currents, which is well-known.  On the other hand, the renormalization of the type (a) operators is nontrivial due to the nonlocality of corresponding operators and potential power divergences.  QCD factorization of both types of operators have been studied in Ref.~\cite{Ma:2017pxb}, in which we found that multiplicative renormalizability of these operators is a necessary condition for the collinear factorization to be valid. 

A lot of efforts have been devoted to explore the UV structure of the type (a) quasi-quark operators \cite{Ishikawa:2016znu,Chen:2016fxx,Monahan:2016bvm,Briceno:2017cpo, Xiong:2017jtn, Constantinou:2017sej, Alexandrou:2017huk, Chen:2017mzz}. 
All-order multiplicative renormalizability of quasi-quark operators has been proved using two different methods: one relies on the auxiliary field technique \cite{Ji:2017oey,Green:2017xeu}, and the other is based on diagrammatic expansion \cite{Ishikawa:2017faj}. These proofs provided a firm theoretical basis for extracting the combination of quark distributions that are not sensitive to gluons from LQCD calculated  hadronic matrix elements of quasi-quark operators \cite{Lin:2014zya,Alexandrou:2015rja,Chen:2016utp,Alexandrou:2016jqi,Zhang:2017bzy,Monahan:2017hpu,Chen:2017lnm,Stewart:2017tvs,Izubuchi:2018srq,Alexandrou:2018pbm,Chen:2018xof,Chen:2018fwa,Liu:2018uuj,Bali:2018spj,Radyushkin:2018nbf,Lin:2018qky,Karpie:2018zaz}.

In this paper, we study the UV divergences of quasi-gluon operators to all orders in QCD perturbation theory. The UV structure of quasi-gluon operators could be much more complicated, as we will explain. We define general bare quasi-gluon operators as
\begin{align}\label{eq:ftgx}
\begin{split}
{\cal O}_{bg}^{\mu \nu \rho\sigma}(\xi) = F^{\mu \nu}(\xi)\,\Phi^{(a)}(\{\xi,0\})\, F^{{\rho\sigma}}(0)\,,
\end{split}
\end{align}
where $\Phi^{(a)}(\xi,0)={\cal P}e^{-ig_s\int_0^{1} \xi\cdot A^{(a)}(r \xi)\,dr}$ is a path ordered gauge link in adjoint representation. To be definite, we assume $\xi_\mu$ along $z$-direction and introduce a unit vector $n^\mu = (0,0,0, 1)$, defining $v\cdot n \equiv v_z$ for any vector $v_\mu$.  Due to the dimensional derivative operator in $F^{\mu\nu}$, superficial power counting tells us that the vertex between gluon field strength and gauge link could be linearly UV divergent. By using a cutoff regularization, one-loop calculation in Refs.~\cite{Wang:2017qyg,Wang:2017eel} indeed shows uncanceled linear divergences for this vertex, which would make the multiplicative renormalization of quasi-gluon operators almost impossible.  For the comparison, the corresponding vertex of quasi-quark operators can be at most logarithmic UV divergent.

Another complication comes from operator mixing via UV renormalization. In principle, the general quasi-gluon operator in Eq.~\eqref{eq:ftgx} could have 36 independent operators after taking into account the antisymmetry of gluon field strength, and all of them could be mixed under renormalization.  In literature, quasi-gluon PDFs are constructed from linear combinations of ${\cal O}_{bg}^{\mu \nu \rho\sigma}$ by contracting them with some ``tensors''. In Refs.~\cite{Ji:2013dva} and \cite{Wang:2017eel} and the first lattice simulation \cite{Fan:2018dxu}, different definitions of quasi-gluon PDFs were obtained by contracting ${\cal O}_{bg}^{\mu \nu \rho\sigma}$ by $n_\mu n_\rho  g_{\nu i}g_\sigma^{i}$, $n_\mu n_\rho  g_{\nu \sigma}$ and $g_{\mu 0}g_{\rho0}  g_{\nu \sigma}$, respectively. Renormalization of these quasi-gluon PDFs is non-trivial due to the mixing of these operators. 

In this paper, we first perform an explicit one-loop calculation of quasi-gluon operators of an asymptotic gluon of momentum $p$, defined as $\langle g(p)|{\cal O}_{bg}^{\mu \nu \rho\sigma}(\xi) | g(p) \rangle$.  We use dimensional regularization (DR) to regularize both logarithmic and linear UV divergences, which respectively appear as poles around $d=4$ and $d=4-1/n$ at $n$-loop order. We find that linear UV divergences of one-loop correction to the gluon-gauge-link vertex are canceled under DR, which makes the multiplicative renormalizability of quasi-gluon operators a possibility. We then explore all possible UV divergent topologies of higher order diagrams.  Using gauge invariance, we find that all linear UV divergences from the gluon-gauge-link vertex are canceled to all-orders in perturbation theory.
Then, we find that all of the 36 independent quasi-gluon operators can be multipliticatively renormalized without mixing with any other operators. Combining with our proof for quasi-quark operators in Ref.~\cite{Ishikawa:2017faj}, our work presented in this paper for quasi-gluon operators completes the proof of multipliticative renormalization of the quasi-parton operators.

\begin{figure}[htb]
\begin{center}
\includegraphics[width=2.8in]{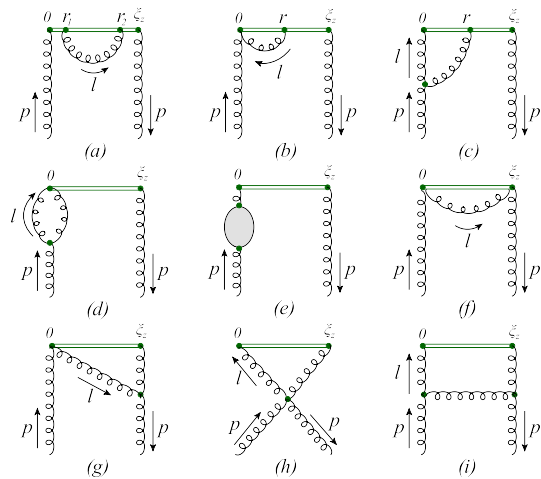}
\caption{\label{fig:oneloop-g}
Some typical Feynman diagrams for quasi-gluon PDFs of an asymptotic gluon of momentum $p$ at one-loop order.}
\end{center}
\end{figure}

%%%%%%%%%%%%%%%%%%%%%%%%%%%%%%%%%%%%%%%%%%%%%%%%%
\sect{UV Divergences at One Loop}
\label{sec:UVOL}
We present the relevant one-loop Feynman diagrams for quasi-gluon operators of an asymptotic gluon of momentum $p$ in Fig.~\ref{fig:oneloop-g}, where the bubble in the diagram (e) includes all one-loop self-energy diagrams of the active gluon.  For the complete one-loop contribution, additional Feynman diagrams are needed.  Some of them are mirror of diagrams (b), (c), (d), (e) and (g), while the rest can be obtained by replacing external momentum $p$ to $- p$ in all these Feynman amplitudes.  For the following one-loop calculation, we take the linearly combined quasi-gluon operator in Ref.~\cite{Ji:2013dva} as an example, but our conclusion is true for any of the 36 independent operators.

%%%%%%%%%%%%%%%%%%%%%%%%%%%%%%%
%%%%%%%%%%%%%%%%%%%%%%%%%%%%%%

We choose Feynman gauge, and assume $\xi_z$ to be positive for definiteness.  The diagram (a) in Fig.~\ref{fig:oneloop-g} gives
\begin{align}
M_{1{a}}&= \frac{g_s^2 \mu_r^{4-d} C_A}{i} \,e^{-i p_z \xi_z} \int_0^{\xi_z}dr_1\int_{r_1}^{\xi_z}dr_2 \frac{d^d l}{(2\pi)^d} \frac{e^{il_z(r_2-r_1)}}{l^2}
\nonumber\\
&\xlongequal{\text{UV}} \frac{\alpha_s C_A}{\pi} e^{-i p_z \xi_z} \left(-\frac{\pi \mu_r\xi_z}{3-d}+ \frac{1}{4-d} \right) \, ,
\label{eq:1a0}
\end{align}
where $\mu_r$ is a renormalization scale to compensate the mass dimension in DR. This diagram contributes to both linear and logarithmic UV divergences, as expected.

To understand where in this one-loop phase space the UV divergences in Eq.~(\ref{eq:1a0}) come from, it is instructive to distinguish $l_z$ - the $z$-component of the loop momentum $l$ from $\bar{l}_\mu$ - the other components of $l$, as $l_\mu = \bar{l}_\mu - l_z n_{\mu}$ with $l^2=\bar{l}^2-l_z^2$ \cite{Ishikawa:2017faj}. 
If $\bar{l}^2$ is constrained in a finite region in Eq.~(\ref{eq:1a0}), integrating $l_z$, $r_1$ and $r_2$ cannot generate any UV divergence.  Furthermore, there is no UV divergence if we do not include the region where $|r_2-r_1|$ is very small, which can be demonstrated by introducing the following decomposition, 
\begin{align}\label{eq:dec}
\frac{1}{(l+q)^2} 
= \frac{1}{(\bar{l}+\bar{q})^2}+ \frac{(l_z+q_z)^2}{(l+q)^2(\bar{l}+\bar{q})^2}\, ,
\end{align}
where $q$ can be any unintegrated momentum. By applying this decomposition to $\frac{1}{l^2} $ in Eq.~\eqref{eq:1a0}, the first term is free of $l_z$, and thus the integration of $l_z$ gives $\delta(r_2-r_1)$, while the second term is UV finite under the integration of $\bar{l}$.  That is, the UV divergence in Eq.~\eqref{eq:1a0} can only come from the region of phase space where $\bar{l}$ are in UV region while $|r_2-r_1|$ is very small. Therefore, we conclude that, with DR, all UV divergences of diagram (a) in Fig.~\ref{fig:oneloop-g} come from a region localized in spacetime. 

By decomposing both $\frac{1}{l^2} $ and $\frac{1}{(p-l)^2}$ using Eq.~\eqref{eq:dec}, we obtain many terms for diagrams (b) and (c) and found that these terms are either free of $l_z$ in denominator, which result in $\delta(r)$ or its derivatives, or UV finite under the integration of $\bar{l}$. Thus the UV divergences of these two diagrams are also localized in spacetime,
\begin{align}
M_{1{b}}
&\xlongequal{\text{UV}} \frac{\alpha_s C_A}{\pi} e^{-i p_z \xi_z} \left(\frac{-i}{p_z \xi_z} \frac{\pi \mu_r \xi_z}{3-d} \right) \, , \label{eq:1b0}\\
M_{1c}
&\xlongequal{\text{UV}} \frac{\alpha_s C_A}{\pi} e^{-i p_z \xi_z} \left(\frac{i}{p_z \xi_z} \frac{\pi \mu_r \xi_z}{3-d}+\frac{3}{4} \frac{1}{4-d} \right) \, ,
\label{eq:1c0} 
\end{align}
where diagram (b) has only linear UV divergence, while (c) has both linear and logarithmic UV divergence.  

The diagrams (d) and (e) in Fig.~\ref{fig:oneloop-g} have only logarithmic UV divergences, which come from the region where all components of $l^\mu$ go to infinity, and thus, is localized in spacetime.
All other diagrams in Fig.~\ref{fig:oneloop-g} are free of UV divergence, simply because the loop cannot be localized in spacetime due to finite $\xi_z$, same as the argument for quasi-quark operators \cite{Ishikawa:2017faj}. 
In summary, we conclude that the UV divergences of quasi-gluon operators at one-loop can only be emerged from a region {\it localized}\ in coordinate spacetime. For a comparison, UV divergences of operators defining PDFs come from a region   {\it nonlocal}\ along the ``$-$" light-cone direction \cite{Ishikawa:2017faj}.

The linear UV divergence in Eq.~(\ref{eq:1a0}) from the diagram (a) is harmless because it can be easily exponentiated to all-order as an overall phase factor and then multiplicatively renormalized, just like the case of quasi-quark operators \cite{Ishikawa:2017faj,Polyakov:1980ca,Dotsenko:1979wb}. However, the presence of linear UV divergence in Eqs.~(\ref{eq:1b0}) and (\ref{eq:1c0}) from the diagrams (b) and (c), respectively, could challenge the multiplicative renormalizability. Fortunately, we find that with DR, the linear UV divergences from these two diagrams are canceled.  On the contrary, the linear divergences from the diagrams (b) and (c) do not cancel if one uses a cutoff regularization that breaks the gauge symmetry \cite{Wang:2017qyg,Wang:2017eel}. This implies that gauge invariance plays an important role to remove the linear UV divergences that may challenge the multiplicative renormalizability.  In the following, we will use gauge invariance to show that all linear divergences, except that from self-energy of gauge links, are canceled by summing over all contributions, and quasi-gluon operators could be multiplicatively renormalized.

\begin{figure}[htb]
	\begin{center}
		\includegraphics[width=2.8in]{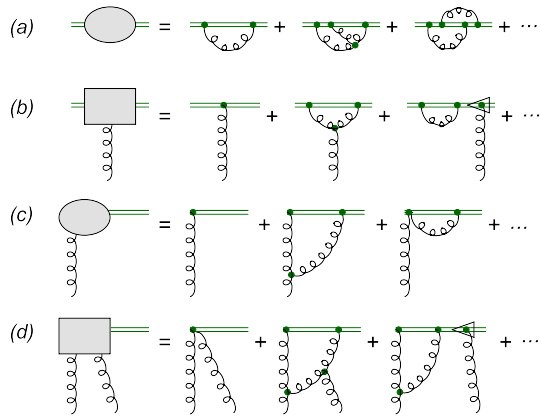}
		\caption{\label{fig:div-topo-g}
			Four topologies of diagrams which could give UV divergences to the quasi-gluon operators.}
	\end{center}
\end{figure}

%%%%%%%%%%%%%%%%%%%%%%%%%%%%%%%%%%%%%%%%%%%%%%%%%
\sect{UV divergences at high orders}
\label{sec:top}
From the one-loop diagrams in Fig.~\ref{fig:oneloop-g}, we can generate all high order loop diagrams by adding gluons, quark-antiquark pairs, or ghost-antighost pairs to them. Because of the isolation of $z$-component in the definition of quasi-parton operators, both 3-dimensional (3-D) and 4-D integration of loop momentum $\bar{l}$ and $l$ could lead to UV divergence.  In Ref.~\cite{Ishikawa:2017faj}, we introduced the change of divergence index $\Delta\omega_3$ and $\Delta\omega_4$ for the 3-D and 4-D integration of loop momenta of higher order diagrams, respectively, and showed that it is sufficient, although it is not necessary, that quasi-parton operators are renormalizable if $\Delta\omega_3\leq 0$ and $\Delta\omega_4\leq 0$ are satisfied for all corresponding higher order diagrams.  Based on the power counting rules derived in Ref.~\cite{Ishikawa:2017faj}, we find that the only case that may increase superficial UV divergence of quasi-gluon operators at high orders is when we add a gluon with both ends of it attached to the gauge link, where the 3-D integration gets $\Delta\omega_3=1$. By applying the decomposition in Eq.~\eqref{eq:dec} to the added gluon's momentum, it is straightforward to show that dimensional regularized UV divergences at any loop level are localized in spacetime, in the same way as quasi-quark operators shown in~\cite{Ishikawa:2017faj}.  As a result, we find that $\Delta\omega_3$ is effectively irrelevant for studying UV divergences. Because $\Delta\omega_4\leq0$ for all cases, there are only finite topologies of high order diagrams in Fig.~\ref{fig:div-topo-g} that have UV divergences.

%%%%%%%%%%%%%%%%%%%%%%%%%%%%%%%
%%%%%%%%%%%%%%%%%%%%%%%%%%%%%%

The blobs with topologies (a) and (c) in Fig.~\ref{fig:div-topo-g} denote one-particle-irreducible diagrams, and they both have linear superficial UV divergences. Because of the potential linear UV divergences, diagrams with one more gluon attached to the blobs can generate logarithmic UV divergences.  Another possibility to produce logarithmic divergences is when a gluon is attached to the gauge link outside of, but very close to, the blobs, as shown in Fig.~\ref{fig:div-topo-g} (b) and (d), with the attachment denoted by a triangle. The blobs of topologies (b) and (d) include both kinds of logarithmic divergent diagrams mentioned here. 

The topologies (a) and (b) in Fig.~\ref{fig:div-topo-g} are the same as that for quasi-quark operators, and their divergences can be renormalized similarly.  Linear divergences from the diagrams of the topology (a) can be removed by an overall factor as the mass renormalization of a test particle moving along the gauge link \cite{Polyakov:1980ca}, and its logarithmic divergences caused by end points of the gauge link can be removed by multiplying $Z_{wg}^{-1/2}$ - the ``wave function" renormalization of the test particle \cite{Dotsenko:1979wb}. The diagrams of the topology (b) has only logarithmic UV divergence, which can be taken care of by QCD  renormalization \cite{Dotsenko:1979wb}.

The UV divergences from diagrams of topologies (c) and (d) in Fig.~\ref{fig:div-topo-g} are different from that of quasi-quark operators, and are studied in next two sections, respectively.

\begin{figure}[htb]
	\begin{center}
		\includegraphics[width=2.in]{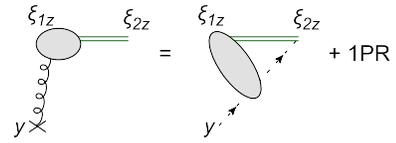}
		\caption{\label{fig:gwi1} Pictorial representation of the generalized Ward identity in Eq.~(\ref{eq:gwi1}) with dashed line represents the ghost field.}
	\end{center}
\end{figure}

%%%%%%%%%%%%%%%%%%%%%%%%%%%%%%%%%%%%%%%%%%%%%%%%%
\sect{Renormalization of gluon-gauge-link vertex}
\label{sec:1g-link}
For the definiteness of following discussion, we assume that the gauge link in diagrams of the topology (c) in Fig.~\ref{fig:div-topo-g} starts at an arbitrary coordinate $\xi_{1z}$ with the operator $F^{\mu \nu}(\xi_{1z})$, and ends at another arbitrary coordinate $\xi_{2z}$ with no additional operators. With the ``bare'' coupling constant $g_s$, and ``bare'' field operators for the gluons, the Faddeev-Popov ghost and the antighost given by the symbols $A$, $c$ and $\bar{c}$ respectively, a generalized Ward identity of the non-Abelian field relevant to this topology can be derived \cite{Collins:1984xc}, 
\begin{align}\label{eq:gwi1}
\begin{split}
\langle -i \partial_\lambda^y A_d^\lambda(y)[\Phi(\{\xi_{2z},\xi_{1z}\})]_{a b} \,  F_{b}^{\mu \nu}(\xi_{1z}) \rangle\\
=\langle g_s \bar{c}_d (y) c_e (\xi_{2z}) [ t_e \Phi(\{\xi_{2z},\xi_{1z}\})]_{a b} \, F_{b}^{\mu \nu}(\xi_{1z}) \rangle ,
\end{split}
\end{align}
where the $t$ represents SU(3) generators of the adjoint representation. 
A pictorial representation of Eq.~(\ref{eq:gwi1}) is given in Fig.~\ref{fig:gwi1}, where ``1PR'' denotes one-particle-reducible diagrams. The topology of the left-hand side of Fig.~\ref{fig:gwi1} is the same as that of the diagram (c) in Fig.~\ref{fig:div-topo-g}, but is contracted with external gluon momentum and expressed in coordinate space. The topology of the first term on the right-hand side of Fig.~\ref{fig:gwi1} is nonlocal in spacetime, and thus has no UV divergence. Furthermore, after the renormalization of QCD Lagrangian and gauge-link-related topologies (a) and (b) in Fig.~\ref{fig:div-topo-g}, UV divergences of 1PR diagrams will be canceled. That is, the generalized Ward identity in Eq.~(\ref{eq:gwi1}) ensures that the topology (c) in Fig.~\ref{fig:div-topo-g} is free of UV divergence if it is contracted by external gluon momentum.

%%%%%%%%%%%%%%%%%%%%%%%%%%%%%%%
%%%%%%%%%%%%%%%%%%%%%%%%%%%%%%

To understand the renormalization of the UV divergence of the topology (c) in Fig.~\ref{fig:div-topo-g}, we represent the diagrams of this topology as $\Gamma^{\lambda \mu \nu}(p,n)$, which could be referred as the gluon-gauge-link vertex, where $p$ and $\lambda$ are the momentum and Lorentz index of the external gluon, respectively. Lorentz symmetry combined with antisymmetry between $\mu$ and $\nu$ enable us to do the general decomposition $\Gamma^{\lambda \mu \nu}(p,n)=\sum_{i=1}^4 c_i \Pi_i^{\lambda \mu \nu}$ with 
\begin{align}
\begin{split}
\Pi_1^{\lambda \mu \nu}&=g^{\mu\lambda} p^\nu-g^{\nu\lambda}p^\mu,\, \Pi_2^{\lambda \mu \nu}=(p^\mu n^\nu-p^\nu n^\mu)n^\lambda,\\
\Pi_3^{\lambda \mu \nu}&=(p^\mu n^\nu-p^\nu n^\mu)p^\lambda, \,\Pi_4^{\lambda \mu \nu}=g^{\mu\lambda} n^\nu-g^{\nu\lambda}n^\mu.
\end{split}
\end{align}
Since $p_\lambda \Gamma^{\lambda \mu \nu}=0$ for the UV divergent terms as discussed above, we obtain $c_2\, p\cdot n + c_3\, p^2 + c_4=0$, and consequently, 
\begin{align}
\begin{split}
\Gamma^{\lambda \mu \nu}(p,n)=&c_1 \Pi_1^{\lambda \mu \nu} + c_2 (\Pi_2^{\lambda \mu \nu}-p\cdot n \Pi_4^{\lambda \mu \nu})\\
&+c_3 (\Pi_3^{\lambda \mu \nu}-p^2 \Pi_4^{\lambda \mu \nu}).
\end{split}
\end{align}
Since $c_1$ and $c_2$ have mass dimension $0$, locality of UV divergences ensures that they can be at most logarithmic divergent, while $c_3$ is UV finite due to its mass dimension at $-1$. The only potential linearly UV divergent coefficient $c_4$ is removed by gauge invariance. We have therefore demonstrated that the cancellation of linear UV divergences of diagrams (b) and (c) in Fig.~\ref{fig:oneloop-g} at one-loop order can be generalized to all orders, which makes the multiplicative renormalizability of quasi-gluon operators a possibility.

At the lowest order in $\alpha_s$, we have $\Gamma^{\lambda \mu \nu}(p,n) \propto \Pi_1^{\lambda \mu \nu}$. If we want $\Gamma^{\lambda \mu \nu}(p,n)$ not to mix with other operators under renormalization, we need its UV divergence to be proportional to $\Pi_1^{\lambda \mu \nu}$ to all orders. Fortunately, it is always true. For the case with $\mu$ (or $\nu$) along $z$-direction or the case with both $\mu$ and $\nu$ not along $z$-direction, the coefficients of $c_2$ are proportional to $\Pi_1^{\lambda \mu \nu}$ or equal to zero, respectively. Therefore, the components of $\Gamma^{\lambda \mu \nu}(p,n)$ do not mix with each others at all, although two different renormalization constants are needed for the two different choices. 

In summary, if we choose either $F^{z \bar{\nu}}$ or $F^{\bar{\mu} \bar{\nu}}$ for gluon-gauge-link vertex, $\Gamma^{\lambda \mu \nu}$, we can remove the UV divergences of the vertex by multiplying a corresponding renormalization factor $Z_{vg1}^{-1/2}$ or $Z_{vg2}^{-1/2}$, respectively.

\begin{figure}[t]
	\begin{center}
		\includegraphics[width=2.in]{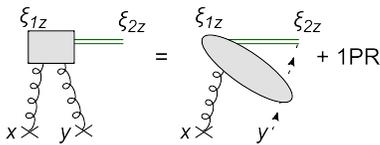}
		\caption{\label{fig:gwi2} Pictorial interpretation of the generalized Ward identity in Eq.~(\ref{eq:gwi2})
		with dashed line represents the ghost field.}
	\end{center}
\end{figure}

%%%%%%%%%%%%%%%%%%%%%%%%%%%%%%%%%%%%%%%%%%%%%%%%%
\sect{Renormalization of gluon-gluon-gauge-link vertex}
Finally, we exam the renormalization of UV divergence of the topology (d) in Fig.~\ref{fig:div-topo-g}, whose diagrams involve two gluons and a gauge link, and could be referred as the gluon-gluon-gauge-link vertex.  Similar to the Ward identity in Eq.~(\ref{eq:gwi1}), we construct the following Ward identity for the ``bare" fields and operators,
\begin{align}
&\langle \partial_\lambda^x A_d^\lambda(x) \partial_\rho^y A_e^\rho(y) [\Phi(\{\xi_{2z},\xi_{1z}\})]_{a b} \,  F_{b}^{\mu \nu}(\xi_{1z})  \rangle \nonumber \\
+&i \delta^{(d)}(x-y) \delta_{de} \langle [\Phi(\{\xi_{2z},\xi_{1z}\})]_{a b} \,  F_{b}^{\mu \nu}(\xi_{1z}) \rangle 
\label{eq:gwi2} \\
=& g_{s} \langle f^{afg} \bar{c}_e (y) c_f (\xi_{2z}) \partial_\lambda^x A_d^\lambda(x) [\Phi(\{\xi_{2z},\xi_{1z}\})]_{g b} \,  F_{b}^{\mu \nu}(\xi_{1z}) \rangle.
\nonumber
\end{align}
A pictorial interpretation of Eq.~(\ref{eq:gwi2}) is given in Fig.~\ref{fig:gwi2}. Similar to Fig.~\ref{fig:gwi1}, the topology of the left-hand side of Fig.~\ref{fig:gwi2} is the same as that of the diagram (d) in Fig.~\ref{fig:div-topo-g}, except the external gluons of the diagrams are contracted with their respective momenta and expressed in coordinate space.  The right-hand side of the equation in Fig.~\ref{fig:gwi2} is UV finite after all previously discussed renormalizations performed, including QCD Lagrangian, gauge links, and gluon-gauge-link vertex. That is,  we find that the topology (d) in Fig.~\ref{fig:div-topo-g} is free of UV divergence if both external gluons are contracted by their respective momenta.  

%%%%%%%%%%%%%%%%%%%%%%%%%%%%%%%
%%%%%%%%%%%%%%%%%%%%%%%%%%%%%%

Similar to the discussion of the gluon-gauge-link vertex, the Ward identity helps reduce the superficial UV divergence of the topology (d) in Fig.~\ref{fig:div-topo-g}. Since the diagrams of the topology (d) have only superficial logarithmic divergence, the additional reduction of the superficial UV divergence from the Ward identity makes the topology (d) in Fig.~\ref{fig:div-topo-g} UV finite. Therefore, after the renormalization of topology (c), the topology (d) requires no  additional renormalization.  This is similar to the case of renormalizing gluon vertexes of QCD Lagrangian, where gauge invariance guarantees that four-gluon vertex will be free of UV divergence once three-gluon vertex is renormalized.

%%%%%%%%%%%%%%%%%%%%%%%%%%%%%%%%%%%%%%%%%%%%%%%%%
\sect{Summary}
We demonstrated that the UV divergences of quasi-gluon operators are actually similar to the UV divergences of quasi-quark operators if all divergences are regularized in a gauge invariant scheme.  We proved that UV divergences of all 36 pure quasi-gluon operators are localized in spacetime, and could be multiplicatively renormalized without mixing with each others,
\begin{align}
{\cal O}_{g}^{\mu \nu \rho\sigma}(\xi)=e^{-C_g |\xi_z|}Z_{wg}^{-1}Z_{vg1}^{-s/2}Z_{vg2}^{-(2-s)/2}{\cal O}_{bg}^{\mu \nu \rho\sigma}(\xi), 
\end{align} 
where $s$ is the number of $z$ components chosen for Lorentz indices $\{ \mu,\nu,\rho,\sigma\}$, and $C_g$, $Z_{wg}$, $Z_{vg1}$ and $Z_{vg2}$ are renormalization constants.
Like the quasi-quark operators, without mixing with each others, hadronic matrix elements of quasi-gluon operators could be examples of good ``lattice cross sections'', as defined in Refs.~\cite{Ma:2014jla,Ma:2017pxb}, which could be calculated in LQCD and factorized into normal PDFs, from which PDFs could be extracted by QCD global analysis of the data of these good ``lattice cross sections'' generated by the first principle LQCD calculations.	

%%%%%%%%%%%%%%%%%%%%%%%%%%%%%%%%%%%%%%%%%%%%%%%%%
\section*{Acknowledgments}
\label{sec:acknowledgments}

We thank Anatoly Radyushkin, George Sterman and Hong Zhang for useful discussions.
This work is supported in part by the U.S. Department of Energy, Office of Science, Office of Nuclear Physics under Award No.~DE-AC05-06OR23177, within the framework of the TMD Topical Collaboration.

{\it Note added:}\ While our paper is being finalized, a preprint by Zhang {\it et al.} \cite{Zhang:2018diq} appeared, in which these authors reached the similar conclusion  although the approach is very different.

%%%%%%%%%%%%%%%%%%%%%%%%%%%%%%%%%%%%%%%%%%%%%%%%%

\providecommand{\href}[2]{#2}\begingroup\raggedright\endgroup

%%%%%%%%%%%%%%%%%%%%%%%%%%%%%%%%%%%%%%%%%%%%%%%%%

\begin{thebibliography}{10}

\bibitem{Ishikawa:2017faj}
T.~Ishikawa, Y.-Q. Ma, J.-W. Qiu, and S.~Yoshida, {\it {Renormalizability of
  quasiparton distribution functions}},
\href{http://dx.doi.org/10.1103/PhysRevD.96.094019}{{\em Phys. Rev.} {\bfseries
  D96} (2017) 094019} [\href{http://arxiv.org/abs/1707.03107}{{\ttfamily
  arXiv:1707.03107}}] [\href{http://inspirehep.net/record/1609459}{{\ttfamily
  InSPIRE}}].
%%CITATION = ARXIV:1707.03107;%%.

\bibitem{Collins:1989gx}
J.~C. Collins, D.~E. Soper, and G.~Sterman, {\it {Factorization of Hard
  Processes in QCD}},
{\em Adv.Ser.Direct.High Energy Phys.} {\bfseries 5} (1988) 1--91
  [\href{http://arxiv.org/abs/hep-ph/0409313}{{\ttfamily hep-ph/0409313}}]
  [\href{http://inspirehep.net/record/25808}{{\ttfamily InSPIRE}}].
%%CITATION = HEP-PH/0409313;%%.

\bibitem{Lin:2017snn}
H.-W. Lin {\em et al.}, {\it {Parton distributions and lattice QCD
  calculations: a community white paper}},
\href{http://dx.doi.org/10.1016/j.ppnp.2018.01.007}{{\em Prog. Part. Nucl.
  Phys.} {\bfseries 100} (2018) 107--160}
  [\href{http://arxiv.org/abs/1711.07916}{{\ttfamily arXiv:1711.07916}}]
  [\href{http://inspirehep.net/record/1637373}{{\ttfamily InSPIRE}}].
%%CITATION = ARXIV:1711.07916;%%.

\bibitem{Ji:2013dva}
X.~Ji, {\it {Parton Physics on a Euclidean Lattice}},
\href{http://dx.doi.org/10.1103/PhysRevLett.110.262002}{{\em Phys.Rev.Lett.}
  {\bfseries 110} (2013) 262002}
  [\href{http://arxiv.org/abs/1305.1539}{{\ttfamily arXiv:1305.1539}}]
  [\href{http://inspirehep.net/record/1232221}{{\ttfamily InSPIRE}}].
%%CITATION = ARXIV:1305.1539;%%.

\bibitem{Liu:1993cv}
K.-F. Liu and S.-J. Dong, {\it {Origin of difference between anti-d and anti-u
  partons in the nucleon}},
\href{http://dx.doi.org/10.1103/PhysRevLett.72.1790}{{\em Phys. Rev. Lett.}
  {\bfseries 72} (1994) 1790--1793}
  [\href{http://arxiv.org/abs/hep-ph/9306299}{{\ttfamily hep-ph/9306299}}]
  [\href{http://inspirehep.net/record/354824}{{\ttfamily InSPIRE}}].
%%CITATION = HEP-PH/9306299;%%.

\bibitem{Liu:1999ak}
K.-F. Liu, {\it {Parton degrees of freedom from the path integral formalism}},
\href{http://dx.doi.org/10.1103/PhysRevD.62.074501}{{\em Phys. Rev.} {\bfseries
  D62} (2000) 074501} [\href{http://arxiv.org/abs/hep-ph/9910306}{{\ttfamily
  hep-ph/9910306}}] [\href{http://inspirehep.net/record/508485}{{\ttfamily
  InSPIRE}}].
%%CITATION = HEP-PH/9910306;%%.

\bibitem{Liu:2016djw}
K.-F. Liu, {\it {Parton Distribution Function from the Hadronic Tensor on the
  Lattice}},
{\em PoS} {\bfseries LATTICE2015} (2016) 115
  [\href{http://arxiv.org/abs/1603.07352}{{\ttfamily arXiv:1603.07352}}]
  [\href{http://inspirehep.net/record/1434313}{{\ttfamily InSPIRE}}].
%%CITATION = ARXIV:1603.07352;%%.

\bibitem{Orginos:2017kos}
K.~Orginos, A.~Radyushkin, J.~Karpie, and S.~Zafeiropoulos, {\it {Lattice QCD
  exploration of parton pseudo-distribution functions}},
\href{http://dx.doi.org/10.1103/PhysRevD.96.094503}{{\em Phys. Rev.} {\bfseries
  D96} (2017) 094503} [\href{http://arxiv.org/abs/1706.05373}{{\ttfamily
  arXiv:1706.05373}}] [\href{http://inspirehep.net/record/1605575}{{\ttfamily
  InSPIRE}}].
%%CITATION = ARXIV:1706.05373;%%.

\bibitem{Chambers:2017dov}
A.~J. Chambers, R.~Horsley, Y.~Nakamura, H.~Perlt, P.~E.~L. Rakow,
  G.~Schierholz, A.~Schiller, K.~Somfleth, R.~D. Young, and J.~M. Zanotti, {\it
  {Nucleon Structure Functions from Operator Product Expansion on the
  Lattice}},
\href{http://dx.doi.org/10.1103/PhysRevLett.118.242001}{{\em Phys. Rev. Lett.}
  {\bfseries 118} (2017) 242001}
  [\href{http://arxiv.org/abs/1703.01153}{{\ttfamily arXiv:1703.01153}}]
  [\href{http://inspirehep.net/record/1516003}{{\ttfamily InSPIRE}}].
%%CITATION = ARXIV:1703.01153;%%.

\bibitem{Ma:2014jla}
Y.-Q. Ma and J.-W. Qiu,
{\it {Extracting Parton Distribution Functions from Lattice QCD Calculations}},
   [\href{http://arxiv.org/abs/1404.6860}{{\ttfamily arXiv:1404.6860}}]
  [\href{http://inspirehep.net/record/1292807}{{\ttfamily InSPIRE}}].
%%CITATION = ARXIV:1404.6860;%%.

\bibitem{Ma:2017pxb}
Y.-Q. Ma and J.-W. Qiu, {\it {Exploring Partonic Structure of Hadrons Using ab
  initio Lattice QCD Calculations}},
\href{http://dx.doi.org/10.1103/PhysRevLett.120.022003}{{\em Phys. Rev. Lett.}
  {\bfseries 120} (2018) 022003}
  [\href{http://arxiv.org/abs/1709.03018}{{\ttfamily arXiv:1709.03018}}]
  [\href{http://inspirehep.net/record/1622746}{{\ttfamily InSPIRE}}].
%%CITATION = ARXIV:1709.03018;%%.

\bibitem{Ishikawa:2016znu}
T.~Ishikawa, Y.-Q. Ma, J.-W. Qiu, and S.~Yoshida,
{\it {Practical quasi parton distribution functions}},
  [\href{http://arxiv.org/abs/1609.02018}{{\ttfamily arXiv:1609.02018}}]
  [\href{http://inspirehep.net/record/1485578}{{\ttfamily InSPIRE}}].
%%CITATION = ARXIV:1609.02018;%%.

\bibitem{Chen:2016fxx}
J.-W. Chen, X.~Ji, and J.-H. Zhang, {\it {Improved quasi parton distribution
  through Wilson line renormalization}},
\href{http://dx.doi.org/10.1016/j.nuclphysb.2016.12.004}{{\em Nucl. Phys.}
  {\bfseries B915} (2017) 1--9}
  [\href{http://arxiv.org/abs/1609.08102}{{\ttfamily arXiv:1609.08102}}]
  [\href{http://inspirehep.net/record/1488113}{{\ttfamily InSPIRE}}].
%%CITATION = ARXIV:1609.08102;%%.

\bibitem{Monahan:2016bvm}
C.~Monahan and K.~Orginos, {\it {Quasi parton distributions and the gradient
  flow}},
\href{http://dx.doi.org/10.1007/JHEP03(2017)116}{{\em JHEP} {\bfseries 03}
  (2017) 116} [\href{http://arxiv.org/abs/1612.01584}{{\ttfamily
  arXiv:1612.01584}}] [\href{http://inspirehep.net/record/1501937}{{\ttfamily
  InSPIRE}}].
%%CITATION = ARXIV:1612.01584;%%.

\bibitem{Briceno:2017cpo}
R.~A. Brice\~{n}o, M.~T. Hansen, and C.~J. Monahan, {\it {Role of the Euclidean
  signature in lattice calculations of quasidistributions and other nonlocal
  matrix elements}},
\href{http://dx.doi.org/10.1103/PhysRevD.96.014502}{{\em Phys. Rev.} {\bfseries
  D96} (2017) 014502} [\href{http://arxiv.org/abs/1703.06072}{{\ttfamily
  arXiv:1703.06072}}] [\href{http://inspirehep.net/record/1518148}{{\ttfamily
  InSPIRE}}].
%%CITATION = ARXIV:1703.06072;%%.

\bibitem{Xiong:2017jtn}
X.~Xiong, T.~Luu, and U.-G. Mei{\ss}ner,
{\it {Quasi-Parton Distribution Function in Lattice Perturbation Theory}},
  [\href{http://arxiv.org/abs/1705.00246}{{\ttfamily arXiv:1705.00246}}]
  [\href{http://inspirehep.net/record/1597417}{{\ttfamily InSPIRE}}].
%%CITATION = ARXIV:1705.00246;%%.

\bibitem{Constantinou:2017sej}
M.~Constantinou and H.~Panagopoulos, {\it {Perturbative renormalization of
  quasi-parton distribution functions}},
\href{http://dx.doi.org/10.1103/PhysRevD.96.054506}{{\em Phys. Rev.} {\bfseries
  D96} (2017) 054506} [\href{http://arxiv.org/abs/1705.11193}{{\ttfamily
  arXiv:1705.11193}}] [\href{http://inspirehep.net/record/1601910}{{\ttfamily
  InSPIRE}}].
%%CITATION = ARXIV:1705.11193;%%.

\bibitem{Alexandrou:2017huk}
C.~Alexandrou, K.~Cichy, M.~Constantinou, K.~Hadjiyiannakou, K.~Jansen,
  H.~Panagopoulos, and F.~Steffens, {\it {A complete non-perturbative
  renormalization prescription for quasi-PDFs}},
\href{http://dx.doi.org/10.1016/j.nuclphysb.2017.08.012}{{\em Nucl. Phys.}
  {\bfseries B923} (2017) 394--415}
  [\href{http://arxiv.org/abs/1706.00265}{{\ttfamily arXiv:1706.00265}}]
  [\href{http://inspirehep.net/record/1602174}{{\ttfamily InSPIRE}}].
%%CITATION = ARXIV:1706.00265;%%.

\bibitem{Chen:2017mzz}
J.-W. Chen, T.~Ishikawa, L.~Jin, H.-W. Lin, Y.-B. Yang, J.-H. Zhang, and
  Y.~Zhao, {\it {Parton Distribution Function with Non-perturbative
  Renormalization from Lattice QCD}},
\href{http://dx.doi.org/10.1103/PhysRevD.97.014505}{{\em Phys. Rev.} {\bfseries
  D97} (2018) 014505} [\href{http://arxiv.org/abs/1706.01295}{{\ttfamily
  arXiv:1706.01295}}] [\href{http://inspirehep.net/record/1602626}{{\ttfamily
  InSPIRE}}].
%%CITATION = ARXIV:1706.01295;%%.

\bibitem{Ji:2017oey}
X.~Ji, J.-H. Zhang, and Y.~Zhao, {\it {Renormalization in Large Momentum
  Effective Theory of Parton Physics}},
\href{http://dx.doi.org/10.1103/PhysRevLett.120.112001}{{\em Phys. Rev. Lett.}
  {\bfseries 120} (2018) 112001}
  [\href{http://arxiv.org/abs/1706.08962}{{\ttfamily arXiv:1706.08962}}]
  [\href{http://inspirehep.net/record/1607797}{{\ttfamily InSPIRE}}].
%%CITATION = ARXIV:1706.08962;%%.

\bibitem{Green:2017xeu}
J.~Green, K.~Jansen, and F.~Steffens, {\it {Nonperturbative renormalization of
  nonlocal quark bilinears for quasi-PDFs on the lattice using an auxiliary
  field}},
\href{http://dx.doi.org/10.1103/PhysRevLett.121.022004}{{\em Phys. Rev. Lett.}
  {\bfseries 121} (2018) 022004}
  [\href{http://arxiv.org/abs/1707.07152}{{\ttfamily arXiv:1707.07152}}]
  [\href{http://inspirehep.net/record/1611305}{{\ttfamily InSPIRE}}].
%%CITATION = ARXIV:1707.07152;%%.

\bibitem{Lin:2014zya}
H.-W. Lin, J.-W. Chen, S.~D. Cohen, and X.~Ji, {\it {Flavor Structure of the
  Nucleon Sea from Lattice QCD}},
\href{http://dx.doi.org/10.1103/PhysRevD.91.054510}{{\em Phys. Rev.} {\bfseries
  D91} (2015) 054510} [\href{http://arxiv.org/abs/1402.1462}{{\ttfamily
  arXiv:1402.1462}}] [\href{http://inspirehep.net/record/1280317}{{\ttfamily
  InSPIRE}}].
%%CITATION = ARXIV:1402.1462;%%.

\bibitem{Alexandrou:2015rja}
C.~Alexandrou, K.~Cichy, V.~Drach, E.~Garcia-Ramos, K.~Hadjiyiannakou,
  K.~Jansen, F.~Steffens, and C.~Wiese, {\it {Lattice calculation of parton
  distributions}},
\href{http://dx.doi.org/10.1103/PhysRevD.92.014502}{{\em Phys. Rev.} {\bfseries
  D92} (2015) 014502} [\href{http://arxiv.org/abs/1504.07455}{{\ttfamily
  arXiv:1504.07455}}] [\href{http://inspirehep.net/record/1365095}{{\ttfamily
  InSPIRE}}].
%%CITATION = ARXIV:1504.07455;%%.

\bibitem{Chen:2016utp}
J.-W. Chen, S.~D. Cohen, X.~Ji, H.-W. Lin, and J.-H. Zhang, {\it {Nucleon
  Helicity and Transversity Parton Distributions from Lattice QCD}},
\href{http://dx.doi.org/10.1016/j.nuclphysb.2016.07.033}{{\em Nucl. Phys.}
  {\bfseries B911} (2016) 246--273}
  [\href{http://arxiv.org/abs/1603.06664}{{\ttfamily arXiv:1603.06664}}]
  [\href{http://inspirehep.net/record/1431989}{{\ttfamily InSPIRE}}].
%%CITATION = ARXIV:1603.06664;%%.

\bibitem{Alexandrou:2016jqi}
C.~Alexandrou, K.~Cichy, M.~Constantinou, K.~Hadjiyiannakou, K.~Jansen,
  F.~Steffens, and C.~Wiese, {\it {Updated Lattice Results for Parton
  Distributions}},
\href{http://dx.doi.org/10.1103/PhysRevD.96.014513}{{\em Phys. Rev.} {\bfseries
  D96} (2017) 014513} [\href{http://arxiv.org/abs/1610.03689}{{\ttfamily
  arXiv:1610.03689}}] [\href{http://inspirehep.net/record/1491383}{{\ttfamily
  InSPIRE}}].
%%CITATION = ARXIV:1610.03689;%%.

\bibitem{Zhang:2017bzy}
J.-H. Zhang, J.-W. Chen, X.~Ji, L.~Jin, and H.-W. Lin, {\it {Pion Distribution
  Amplitude from Lattice QCD}},
\href{http://dx.doi.org/10.1103/PhysRevD.95.094514}{{\em Phys. Rev.} {\bfseries
  D95} (2017) 094514} [\href{http://arxiv.org/abs/1702.00008}{{\ttfamily
  arXiv:1702.00008}}] [\href{http://inspirehep.net/record/1511678}{{\ttfamily
  InSPIRE}}].
%%CITATION = ARXIV:1702.00008;%%.

\bibitem{Monahan:2017hpu}
C.~Monahan, {\it {Smeared quasidistributions in perturbation theory}},
\href{http://dx.doi.org/10.1103/PhysRevD.97.054507}{{\em Phys. Rev.} {\bfseries
  D97} (2018) 054507} [\href{http://arxiv.org/abs/1710.04607}{{\ttfamily
  arXiv:1710.04607}}] [\href{http://inspirehep.net/record/1630477}{{\ttfamily
  InSPIRE}}].
%%CITATION = ARXIV:1710.04607;%%.

\bibitem{Chen:2017lnm}
J.-W. Chen, T.~Ishikawa, L.~Jin, H.-W. Lin, A.~Sch\"afer, Y.-B. Yang, J.-H.
  Zhang, and Y.~Zhao,
{\it {Gaussian-weighted Parton Quasi-distribution}},
  [\href{http://arxiv.org/abs/1711.07858}{{\ttfamily arXiv:1711.07858}}]
  [\href{http://inspirehep.net/record/1637383}{{\ttfamily InSPIRE}}].
%%CITATION = ARXIV:1711.07858;%%.

\bibitem{Stewart:2017tvs}
I.~W. Stewart and Y.~Zhao, {\it {Matching the quasiparton distribution in a
  momentum subtraction scheme}},
\href{http://dx.doi.org/10.1103/PhysRevD.97.054512}{{\em Phys. Rev.} {\bfseries
  D97} (2018) 054512} [\href{http://arxiv.org/abs/1709.04933}{{\ttfamily
  arXiv:1709.04933}}] [\href{http://inspirehep.net/record/1623909}{{\ttfamily
  InSPIRE}}].
%%CITATION = ARXIV:1709.04933;%%.

\bibitem{Izubuchi:2018srq}
T.~Izubuchi, X.~Ji, L.~Jin, I.~W. Stewart, and Y.~Zhao,
{\it {Factorization Theorem Relating Euclidean and Light-Cone Parton
  Distributions}},  [\href{http://arxiv.org/abs/1801.03917}{{\ttfamily
  arXiv:1801.03917}}] [\href{http://inspirehep.net/record/1647571}{{\ttfamily
  InSPIRE}}].
%%CITATION = ARXIV:1801.03917;%%.

\bibitem{Alexandrou:2018pbm}
C.~Alexandrou, K.~Cichy, M.~Constantinou, K.~Jansen, A.~Scapellato, and
  F.~Steffens,
{\it {Reconstruction of light-cone parton distribution functions from lattice
  QCD simulations at the physical point}},
  [\href{http://arxiv.org/abs/1803.02685}{{\ttfamily arXiv:1803.02685}}]
  [\href{http://inspirehep.net/record/1658890}{{\ttfamily InSPIRE}}].
%%CITATION = ARXIV:1803.02685;%%.

\bibitem{Chen:2018xof}
J.-W. Chen, L.~Jin, H.-W. Lin, Y.-S. Liu, Y.-B. Yang, J.-H. Zhang, and Y.~Zhao,
{\it {Lattice Calculation of Parton Distribution Function from LaMET at
  Physical Pion Mass with Large Nucleon Momentum}},
  [\href{http://arxiv.org/abs/1803.04393}{{\ttfamily arXiv:1803.04393}}]
  [\href{http://inspirehep.net/record/1662058}{{\ttfamily InSPIRE}}].
%%CITATION = ARXIV:1803.04393;%%.

\bibitem{Chen:2018fwa}
J.-W. Chen, L.~Jin, H.-W. Lin, Y.-S. Liu, A.~Sch\"afer, Y.-B. Yang, J.-H.
  Zhang, and Y.~Zhao,
{\it {First direct lattice-QCD calculation of the $x$-dependence of the pion
  parton distribution function}},
  [\href{http://arxiv.org/abs/1804.01483}{{\ttfamily arXiv:1804.01483}}]
  [\href{http://inspirehep.net/record/1665821}{{\ttfamily InSPIRE}}].
%%CITATION = ARXIV:1804.01483;%%.

\bibitem{Liu:2018uuj}
Y.-S. Liu, J.-W. Chen, L.~Jin, H.-W. Lin, Y.-B. Yang, J.-H. Zhang, and Y.~Zhao,
{\it {Unpolarized quark distribution from lattice QCD: A systematic analysis of
  renormalization and matching}},
  [\href{http://arxiv.org/abs/1807.06566}{{\ttfamily arXiv:1807.06566}}]
  [\href{http://inspirehep.net/record/1682768}{{\ttfamily InSPIRE}}].
%%CITATION = ARXIV:1807.06566;%%.

\bibitem{Bali:2018spj}
{Bali, Gunnar S. and Braun, Vladimir M. and Gl\"a{\ss}le, Benjamin and
  G\"ockeler, Meinulf and Gruber, Michael and Hutzler, Fabian and Korcyl, Piotr
  and Sch\"afer, Andreas and Wein, Philipp and Zhang, Jian-Hui},
{\it {Pion distribution amplitude from Euclidean correlation functions:
  Exploring universality and higher twist effects}},
  [\href{http://arxiv.org/abs/1807.06671}{{\ttfamily arXiv:1807.06671}}]
  [\href{http://inspirehep.net/record/1682987}{{\ttfamily InSPIRE}}].
%%CITATION = ARXIV:1807.06671;%%.

\bibitem{Radyushkin:2018nbf}
A.~V. Radyushkin,
{\it {Structure of parton quasi-distributions and their moments}},
  [\href{http://arxiv.org/abs/1807.07509}{{\ttfamily arXiv:1807.07509}}]
  [\href{http://inspirehep.net/record/1683112}{{\ttfamily InSPIRE}}].
%%CITATION = ARXIV:1807.07509;%%.

\bibitem{Lin:2018qky}
H.-W. Lin, J.-W. Chen, L.~Jin, Y.-S. Liu, Y.-B. Yang, J.-H. Zhang, and Y.~Zhao,
{\it {Spin-Dependent Parton Distribution Function with Large Momentum at
  Physical Pion Mass}},  [\href{http://arxiv.org/abs/1807.07431}{{\ttfamily
  arXiv:1807.07431}}] [\href{http://inspirehep.net/record/1683262}{{\ttfamily
  InSPIRE}}].
%%CITATION = ARXIV:1807.07431;%%.

\bibitem{Karpie:2018zaz}
J.~Karpie, K.~Orginos, and S.~Zafeiropoulos,
{\it {Moments of Ioffe time parton distribution functions from non-local matrix
  elements}},  [\href{http://arxiv.org/abs/1807.10933}{{\ttfamily
  arXiv:1807.10933}}] [\href{http://inspirehep.net/record/1684322}{{\ttfamily
  InSPIRE}}].
%%CITATION = ARXIV:1807.10933;%%.

\bibitem{Wang:2017qyg}
W.~Wang, S.~Zhao, and R.~Zhu, {\it {Gluon quasidistribution function at one
  loop}},
\href{http://dx.doi.org/10.1140/epjc/s10052-018-5617-3}{{\em Eur. Phys. J.}
  {\bfseries C78} (2018) 147}
  [\href{http://arxiv.org/abs/1708.02458}{{\ttfamily arXiv:1708.02458}}]
  [\href{http://inspirehep.net/record/1614777}{{\ttfamily InSPIRE}}].
%%CITATION = ARXIV:1708.02458;%%.

\bibitem{Wang:2017eel}
W.~Wang and S.~Zhao, {\it {On the power divergence in quasi gluon distribution
  function}},
\href{http://dx.doi.org/10.1007/JHEP05(2018)142}{{\em JHEP} {\bfseries 05}
  (2018) 142} [\href{http://arxiv.org/abs/1712.09247}{{\ttfamily
  arXiv:1712.09247}}] [\href{http://inspirehep.net/record/1644909}{{\ttfamily
  InSPIRE}}].
%%CITATION = ARXIV:1712.09247;%%.

\bibitem{Fan:2018dxu}
Z.-Y. Fan, Y.-B. Yang, A.~Anthony, H.-W. Lin, and K.-F. Liu,
{\it {Gluon Quasi-PDF From Lattice QCD}},
  [\href{http://arxiv.org/abs/1808.02077}{{\ttfamily arXiv:1808.02077}}]
  [\href{http://inspirehep.net/record/1685340}{{\ttfamily InSPIRE}}].
%%CITATION = ARXIV:1808.02077;%%.

\bibitem{Polyakov:1980ca}
A.~M. Polyakov, {\it {Gauge Fields as Rings of Glue}},
\href{http://dx.doi.org/10.1016/0550-3213(80)90507-6}{{\em Nucl. Phys.}
  {\bfseries B164} (1980) 171--188}
  [\href{http://inspirehep.net/record/157352}{{\ttfamily InSPIRE}}].
%%CITATION = NUPHA,B164,171;%%.

\bibitem{Dotsenko:1979wb}
V.~Dotsenko and S.~Vergeles, {\it {Renormalizability of Phase Factors in the
  Nonabelian Gauge Theory}},
\href{http://dx.doi.org/10.1016/0550-3213(80)90103-0}{{\em Nucl.Phys.}
  {\bfseries B169} (1980) 527}
  [\href{http://inspirehep.net/record/144525}{{\ttfamily InSPIRE}}].
%%CITATION = NUPHA,B169,527;%%.

\bibitem{Collins:1984xc}
J.~C. Collins, {\em {Renormalization. An Introduction To Renormalization, The
  Renormalization Group, And The Operator Product Expansion}}.
\newblock
\newblock
1984 [\href{http://inspirehep.net/record/209810}{{\ttfamily InSPIRE}}].
\newblock
%%CITATION = INSPIRE-209810;%%.

\bibitem{Zhang:2018diq}
J.-H. Zhang, X.~Ji, A.~Sch\"afer, W.~Wang, and S.~Zhao,
{\it {Renormalization of gluon quasi-PDF in large momentum effective theory}},
  [\href{http://arxiv.org/abs/1808.10824}{{\ttfamily arXiv:1808.10824}}]
  [\href{http://inspirehep.net/record/1692401}{{\ttfamily InSPIRE}}].
%%CITATION = ARXIV:1808.10824;%%.

\end{thebibliography}
\end{document}